\documentclass{webofc}
\usepackage[varg]{txfonts}   
\usepackage{pifont}
\newcommand{\mytexttilde}{{\raise.17ex\hbox{$\scriptstyle\mathtt{\sim}$}}}

\def\amu{a_\mu}
\def\amuh{a_\mu^{{\mathrm{had}}}}

\def\mz{M_Z^2}
\def\MZ{M_Z}

\def\dah{\Delta\alpha^{(5)}_{\rm had}}
\def\dahs{\Delta\alpha^{(5)}_{\rm had}(s)}

\def\dah0{\Delta\alpha^{(5)}_{\rm had}(-s_0)}

\def\dahm0{\Delta\alpha^{(5)}_{\rm had}(-M^2_0)}

\newcommand{\gv}{\mbox{GeV}}

\newcommand{\mbo}[1]{$#1$ }
\newcommand{\epm}{e^+e^- }

\newcommand{\power}[1]{\times 10^{#1} }

\newcommand{\D}{\rm d }
\newcommand{\E}{\rm e }

\newcommand{\semis}{\;;\;\; }
\newcommand{\bea}{\begin{eqnarray*}}
\newcommand{\eea}{\end{eqnarray*}}
\newcommand{\ba}{\begin{eqnarray}}
\newcommand{\ea}{\end{eqnarray}}
\newcommand{\epo}{\;. }

\newcommand{\braket}[1]{\langle{#1}\rangle}

\newcommand{\bary}{\begin{array}}
\newcommand{\eary}{\end{array}}

\newcommand{\SU}{\mathit{SU}}
\newcommand{\be}{\begin{equation}}
\newcommand{\ee}{\end{equation}}

\renewcommand{\braket}[1]{\langle{#1}\rangle}

\newcommand{\labelitemv}{{\ding{52}} }
\newcommand{\labelitemx}{{\ding{56}} }

\begin{document}
\title{Variations on Photon Vacuum Polarization}
%
%

\author{\firstname{Fred} \lastname{Jegerlehner}\inst{1,2}\fnsep\thanks{\email{fjeger@physik.hu-berlin.de
    }
}
}
\institute{Deutsches Elektronen-Synchrotron (DESY), Platanenallee 6, D-15738 Zeuthen, Germany
\and
          Humboldt-Universit\"at zu Berlin, Institut f\"ur Physik,
       Newtonstrasse 15, D-12489 Berlin, Germany
          }

\abstract{%
I provide updates for the theoretical predictions of the muon and
electron anomalous magnetic moments, for the shift in the fine
structure constant $\alpha(\MZ)$ and for the weak mixing parameter $\sin^2
\Theta_W(M_Z)$. Phenomenological results for Euclidean time correlators,
the key objects in the lattice QCD approach to hadronic vacuum
polarization, are briefly considered. Furthermore, I present a list of
isospin breaking and electromagnetic corrections for the lepton
moments, which may be used to supplement lattice QCD results obtained
in the isospin limit and without the e.m. corrections.}
%
\thispagestyle{empty}
\begin{flushright}
DESY 17-194,~~HU-EP-17/24\\
November 2017
\end{flushright}

\vfill

\begin{center}
{\large\bf
Variations on Photon Vacuum Polarization}\\
{Fred Jegerlehner}\\
{Deutsches  Elektronen--Synchrotron (DESY), Platanenallee 6,\\ D--15738 Zeuthen, Germany\\
Humboldt--Universit\"at zu Berlin, Institut f\"ur Physik, Newtonstrasse 15,\\ D--12489 Berlin,
Germany}

\vfill

\begin{minipage}{0.8\textwidth}
{\bf Abstract}\\
I provide updates for the theoretical predictions of the muon and
electron anomalous magnetic moments, for the shift in the fine
structure constant $\alpha(\MZ)$ and for the weak mixing parameter $\sin^2
\Theta_W(M_Z)$. Phenomenological results for Euclidean time correlators,
the key objects in the lattice QCD approach to hadronic vacuum
polarization, are briefly considered. Furthermore, I present a list of
isospin breaking and electromagnetic corrections for the lepton
moments, which may be used to supplement lattice QCD results obtained
in the isospin limit and without the e.m. corrections.
\end{minipage}
\end{center}
\vfill
\noindent\rule{8cm}{0.5pt}\\
$^*$ Invited talk
\textit{International Workshop on $e^+e^-$ collisions from Phi to Psi 2017},
26-29 June 2017, Mainz, Germany
\setcounter{page}{0}
\newpage

\maketitle
\section{Introduction}
\label{intro}
I present some supplementary material on hadronic vacuum polarization
effects which had not been included in my recent
book~\cite{Jegerlehner:2017gek} and the Frascati and Capri
proceedings~\cite{Jegerlehner:2017lbd,Jegerlehner:2015stw}. On the
data side recent BaBar exclusive channel data, BES-III, KEDR, CMD-3
and SND data are actualized (see these proceedings). Besides
continuous progress in $\epm$ data also lattice QCD (LQCD) is coming
closer and actually has provided results not available from
elsewhere. This concerns information required for the evaluation of
the $\SU2$ gauge coupling $\alpha_2(s)$ which together with
$\alpha(s)$ allows us to calculate the running weak mixing parameter
$\sin^2 \Theta_W(s)$.
A comparison with lattice results allows one to check the right
strategy of the required flavor recombination.

In view of the upcoming new muon $g-2$ experiments~\cite{LeeRoberts,Mibe}
still the biggest challenge are improved $\epm$ hadronic cross section
measurements for improving hadronic vacuum polarization and $\gamma
\gamma \to \mathrm{ \ hadrons}$ related cross section data for 
improving the hadronic light-by-light contribution. Substantial
progress in lattice QCD calculations of the hadronic current
correlators more and more produce important results which complement
the dispersive approaches~\cite{Pauk,Colangelo}.

\section{HVP for the muon anomaly}
The present status for the hadronic and weak contributions may be
summarized by 
\begin{eqnarray}
\begin{tabular}{lccc}
$ a_\mu^{\mathrm{had}(1)}$&=&$(689.46\pm 3.25)[688.77\pm 3.38][688.07\pm1.14]\:10^{-10}$ &(LO) \\
$ a_\mu^{\mathrm{had}(2)}$&=&$(-99.27\pm 0.67)\:10^{-10}$ & (NLO)\\
$ a_\mu^{\mathrm{had}(3)}$&=&$(1.224\pm 0.010)\:10^{-10}$ & (NNLO)~\cite{Kurz:2014wya}\\
$ a_\mu^\mathrm{had, LbL}$&=&$(10.34\pm 2.88) \power{-10}$ & (HLbL)\\
$ a_\mu^{\mathrm{weak}}  $&=&$(15.36\pm 0.11[m_H,m_t]\pm 0.023
[\rm had])\:10^{-10}$ & (LO+NLO)\epo
\end{tabular}
\end{eqnarray}
For details I refer
to~\cite{Jegerlehner:2017gek,Jegerlehner:2017lbd,Jegerlehner:2015stw}
and references therein (see also~\cite{Zhang:2015yfi,Hagiwara:2017zod}). The QED
prediction of $a_\mu$ is given by (see~\cite{Aoyama:2012wj,Laporta:2017okg,Steinhauser})
\begin{eqnarray}
a_\mu^{\rm QED}
 &=& \frac{\alpha}{ 2\pi}
 +0.765\,857\,423(16) \left( \frac{\alpha }{ \pi}\right)^2\nonumber\\
&& \hspace*{-2cm}
 +24.050\,509\,82(28) \left( \frac{\alpha }{ \pi}\right)^3
+130.8734(60)\left( \frac{\alpha }{ \pi}\right)^4
+751.917(932)\left(\frac{\alpha }{ \pi}\right)^5.
\end{eqnarray}
Given the CODATA/PDG recommended value of $\alpha$ the theory
confronts experiment as collected in Table~\ref{tab:thevsexp}. 
{\small
\begin{table}[h]
\caption{Standard model theory and experiment
comparison}
\label{tab:thevsexp}
\centering
\begin{tabular}{lr@{.}lr@{.}lc}
\hline
Contribution & \multicolumn{2}{c}{Value$\times 10^{10}$} & \multicolumn{2}{l}{Error$\times 10^{10}$} & Reference \\
\hline
QED incl. 4-loops + 5-loops & 11\,658\,471&886 & 0&003 &
\cite{Aoyama:2012wj,Laporta:2017okg}  \\
Hadronic LO vacuum polarization & 689&46 &  3&25 &  \\
Hadronic light--by--light &   10&34 & 2&88 & \\
Hadronic HO vacuum polarization & -8&70 & 0&06 &  \\
Weak to 2-loops & 15&36 & 0&11 & \cite{Gnendiger:2013pva}  \\
\hline Theory & 11\,659\,178&3 & 3&5 & --  \\
Experiment & 11\,659\,209&1 & 6&3 & \cite{BNL04}  \\
The. - Exp.   4.3 standard deviations & -30&6 & 7&2 & -- \\ \hline
\end{tabular}
\end{table}}
As is well known a ``New Physics'' interpretation of the persisting 3
to 4 $\sigma$ difference between prediction and experiment requires
relatively strongly coupled states in the range below about 250
GeV. Search bounds from LEP, Tevatron and specifically from the LHC
already have ruled out a variety of Beyond the Standard Model (BSM)
scenarios, so much hat standard motivations of SUSY/GUT extensions
seem to fall in disgrace.

There is no doubt that performing doable improvements on both the
theory and the experimental side allows one to substantially sharpen
(or diminish) the apparent gap between theory and experiment. Yet,
even the present situation gives ample reason for
speculations. Besides the proton
radius puzzle (PRP)~\cite{Krauth:2017ijq}, no other experimental result has as
many problems to be understood in terms of SM physics. {\footnotesize \textit{Note
added:} The PRP has been solved by now~\cite{PRP}. A new very precise determination of the Rydberg
constant in hydrogen spectroscopy reveals a value by 3 $\sigma$'s
lower than the 2014 CODATA value.}

In any case $\amu$  constrains BSM scenarios distinctively and
at the same time challenges a better understanding of the SM prediction.

\section{ HVP for the electron anomaly}
For the electron anomaly the hadronic and weak contributions read 
\begin{eqnarray}
\begin{tabular}{lccc}
$ a_e^{\mathrm{had}(1)}   $&=&$(184.90\pm 1.08)\:10^{-14}$ &(LO) \\
$ a_e^{\mathrm{had}(2)}   $&=&$(-22.13\pm 0.12)\:10^{-14}$ & (NLO)\\
$ a_e^{\mathrm{had}(3)}   $&=&$(2.80\pm 0.02)\:10^{-14}$ &(NNLO)~\cite{Kurz:2014wya}\\
$ a_e^{\mathrm{had,~ LbL}}$&=&$(3.7\pm 0.5) \power{-14}$ & (HLbL)\\
$ a_e^{\mathrm{weak}}     $&=&$(3.053\pm 0.002[m_H,m_t]\pm 0.023 [\rm had])\:10^{-14}$ & (LO+NLO) \epo
\end{tabular}
\end{eqnarray}
The QED prediction of $a_e$ including the recent results~\cite{Aoyama:2012wj,Laporta:2017okg}
is given by
\begin{eqnarray}
a_e^{\rm QED}
 &=& \frac{\alpha }{ 2\pi}
 -0.328\,478\,444\,002\,54(33) \left( \frac{\alpha }{ \pi}\right)^2\nonumber\\
&&
 +1.181\, 234\, 016\, 816(11) \left( \frac{\alpha }{ \pi}\right)^3
-1.91134(182)\left( \frac{\alpha }{ \pi}\right)^4
+7.791(580)\left(\frac{\alpha }{ \pi}\right)^5.
\end{eqnarray}
The new quasi--analytic $O(\alpha^4)$ result by
Laporta~\cite{Laporta:2017okg} is certainly a milestone in
consolidating the QED part $a_e^{\rm QED}$. For extracting
$ \alpha_{\rm QED}$ the SM prediction
\be
a_e^{\rm SM} = a_e^{\rm QED}
+1.723(12)\times 10^{-12}
\mbox{(hadronic \& weak)}
\ee
is to be confronted with $a_e^\mathrm{exp}=1159\,652\,180.73(28)$ from
experiment~\cite{aenew} as an input. I obtain
\mbo{\alpha^{-1}(a_e)=137.035\,999\,1550(331)(0)(27)(14)[333]\epo}
Using $\alpha$ from atomic interferometry, specifically
$\alpha(h/M_{\rm Rb11})[0.66\, \mathrm{ppb}]$
[$\alpha^{-1}=137.035999037(91)$], the prediction of $a_e$, in units
$10^{-12}$, reads $a_e^\mathrm{the}=1159\,652\,177.28(77)(0)(4)$
[universal] + $2.738(0)$ [$\mu $--loops] + $0.009(0)$ [$\tau$--loops]
+ $1.693(13) $ [hadronic] + $0.030(0)$ [weak] =
$1159\,652\,181.73(77)$ from SM theory, which confronts
$a_e^\mathrm{exp}$.  Thus
\be
a_e^\mathrm{exp}-a_e^\mathrm{the}=-1.00(0.82) \times 10^{-12}\,,
\label{aeexpthe}
\ee
theory and experiment are in excellent agreement. We know that the
sensitivity to new physics is reduced by $(m_\mu/m_e)^2\cdot \delta
a^{\rm exp}_e/\delta a^{\rm exp}_\mu\simeq 19$ relative to
$a_\mu$. Nevertheless, one has to keep in mind that $a_e$ is suffering
less from hadronic uncertainties and thus may provide a safer
test. One should also keep in mind that experiments determining $a_e$
on the one hand and $a_\mu$ on the other hand are very different with
different systematics. While $a_e$ is determined in a ultra cold
environment $a_\mu$ has been determined with ultra relativistic (magic
$\gamma$) muons so far. Presently, the $a_e$ prediction is limited by
the, by a factor $\delta \alpha(\mathrm{Rb11})/\delta
\alpha(a_e)\simeq 5.3$ less precise, $\alpha$ available. Combining all
uncertainties $\amu$ is about a factor 43 more sensitive to new
physics at present.

\section{Hadronic VP and $\alpha(M^2_Z)$}
\label{sec-1}
The running electromagnetic fine structure constant is given by
$\alpha(s)=\alpha/(1-\Delta
\alpha (s))$ with  $\Delta \alpha(s)=\Delta \alpha_{\rm lep}(s)+\dahs +\Delta \alpha_{\rm top}(s)$
where the non-perturbative part evaluated in terms of $\epm$ data reads
\be
\Delta \alpha_{\rm hadrons}^{(5)}(\mz) = 0.027738 \pm 0.000158\,[0.027523 \pm 0.000119]\,,
\ee
where the second result has been obtained with the Euclidean split
technique (Adler function approach). The related $\alpha$ then
corresponds to
\be 
\alpha^{-1}(\mz) = 128.919 \pm 0.022\,[128.958 \pm 0.016] \epo
\ee
Reducing uncertainties via the Euclidean split technique works as
follows: one may split the calculation as
\be
\alpha(\mz)=\alpha^{\mathrm{data}}(-s_0)+\left[\alpha(-\mz)-\alpha(-s_0)\right]^{\mathrm{pQCD}}+\left[\alpha(\mz)-\alpha(-\mz)\right]^{\mathrm{pQCD}}\,,
\ee
where the space-like \mbo{-s_0} is chosen such that pQCD is well under
control in the deep Euclidean region \mbo{-s<-s_0}. The monitor to
control the applicability of pQCD is the Adler function
$D(Q^2)$~\cite{EJKV98}. It reveals that in the space-like region
pQCD works well to predict $D(Q^2)$ down to $s_0= (2.0\,\gv )^2$.
We then may safely use $D^\mathrm{pQCD}(Q^2)$ to calculate perturbatively
\ba
\alpha(-\mz)-\alpha(-s_0)= \frac{\alpha}{3\pi} \int\limits_{s_0}^{\mz}\,
\D Q^{'2} \frac{D^\mathrm{pQCD}(Q^{'2})}{Q^{'2}}\semis
D(Q^2=-s)
=-(12\pi^2)\,s\,\frac{\D
\Pi'_\gamma (s)}{\D s}\epo
\ea
For the offset $s_0=(2.0\, \gv)^2$ I
obtain~\cite{FJ98,Jegerlehner:2008rs} $\Delta\alpha^{(5)}_{\rm
had}(-s_0)^{\mathrm{data}} = 0.006409 \pm 0.000063$,
$\Delta\alpha^{(5)}_{\rm had}(-M_Z^2) = 0.027483 \pm 0.000118$,
$\Delta\alpha^{(5)}_{\rm had}( M_Z^2) = 0.027523 \pm 0.000119$. A
shift $+0.000008$ from the 5-loop contribution is included and an
error $\pm 0.000100$ has been added in quadrature form the perturbative
part. The QCD parameters used are $\alpha_s(M_Z)=0.1189(20)$,
$m_c(m_c)=1.286(13)~[M_c=1.666(17)]~\gv\,,$
$m_b(m_c)=4.164(25)~[M_b=4.800(29)]~\gv\,$, and the evaluation is
based on a complete 3--loop massive QCD analysis~\cite{Chetyrkin:1996cf,Chetyrkin:1997qi}.
Note: the Adler function monitored space-like data vs pQCD split
approach is only moderately more pQCD-driven than the time-like
approach adopted
in~\cite{Davier:2003pw,Ghozzi:2003yn,Davier:2009ag,Zhang:2015yfi}
and by others. For the first direct measurements of $\dahs$ in the
$\rho$ resonance region see~\cite{KLOE-2:2016mgi}.

\section{Hadronic VP and $\alpha_2(M^2_Z)$}
\begin{figure}[h]
\centering
\includegraphics[height=7.6cm]{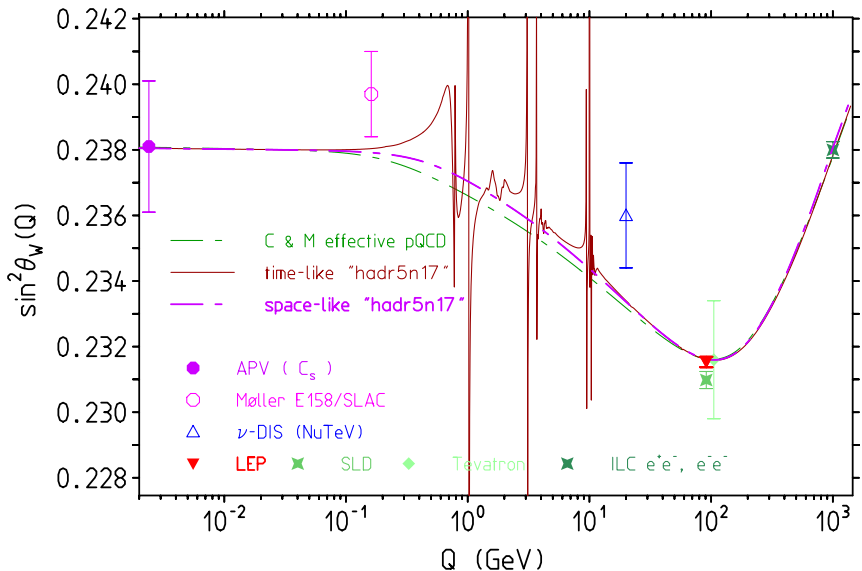}
\caption{$\sin^2 \Theta_W(Q)$ as a function of $Q$ in the time-like
and space-like region. Hadronic uncertainties are included but barely
visible in this plot. Uncertainties from the input parameter $\sin^2
\theta_W(0)=0.23822(100)$ or $\sin^2 \theta_W(M_Z^2)=0.23153(16)$ are
not shown. Note the substantial difference from applying pQCD with
effective quark masses. Future ILC/FCC measurements at 1 TeV would be sensitive to
$Z'$, $H^{--}$ etc.}
\label{fig:sin2theta}
\end{figure}

In electroweak precision physics non-perturbative hadronic effect
primarily show up via the gauge boson self-energy functions. A
prominent example is the scale dependence of the weak mixing parameter
$\sin^2 \Theta_W(s)$. Note that \mbo{\sin^2 \Theta_W(0)/\sin^2
\Theta_W(M_Z^2)=1.02876} a 3\% correction established at
$6.5~\sigma$. To understand this one needs precise information of the
$\SU(2)$ running gauge coupling $\alpha_2(s)$. The hadronic shift is
related to the correlator $\braket{3\gamma}$ where ``3'' marks the
3$^{\rm rd}$ component of the weak isospin  current and ``$\gamma$'' the
e.m. current. As in the case of
$\alpha(s)$ the non-perturbative hadronic contribution can be
evaluated in terms of $\epm$ data in conjunction with separating and
rewighting the various flavor
contributions~\cite{Jegerlehner:1985gq,Jegerlehner:2011mw}. This has
been implemented in the 2016/17 versions of the {\tt alphaQED}
package~\cite{alphaQED17}. The changes affect the $\alpha_2(s)$
routines {\tt alpha2SMr17.f, alpha2SMc17.f} and the $\sin^2
\theta_{\rm eff}$ routine {\tt ACWMsin2theta.f}. The different trials
are compared in Tab.~\ref{tab:flavor} and the updated $\sin^2
\Theta_W(s)$ is shown in Fig.~\ref{fig:sin2theta} for time-like as
well as for space-like momentum transfer. 
Except from the LEP and SLD points (which deviate by {1.8 $\sigma$}), all existing measurements are of rather
limited accuracy unfortunately! Upcoming experiments will improve
results at low space-like $Q$ substantially. 

\section{Euclidean correlators testing flavor separation and reweighting}
Here, we consider the calculation of Euclidean time correlators, which can be
calculated in lattice QCD~\cite{Meyer:2011um,Bernecker:2011gh}. The
aim is to compare lattice results with evaluations obtainable from the
data. As we know, in the low energy region assuming $\SU(3)$ flavor
symmetry is not a good approximation. 
The $\SU(2)$ version assuming OZI violating effects to be negligible
corresponds to a \textbf{perturbative reweighting}! This has been
implemented in the 2012 version of the {\tt alphaQED} package.  Later,
lattice evaluations \cite{Francis:2013jfa,Burger:2015lqa} revealed
this to mismatch the data, while the ``old'' \cite{Jegerlehner:1985gq}
agreed much better, see
\cite{Burger:2015lqa}. Nevertheless, the $\SU(3)$ flavor symmetry
argument also looks the be rather crude when looking at correlator in
the low energy regime. In place of the untenable assumption that OZI
violating terms are small, we may argue by isovector $\rho$ meson
dominance (VMD isovector) which suggests an isospin factor 1/2 in
place of 9/20 suggested by perturbative reweighting. A 10\% difference
in the $ud$ part.

Besides the flavor $\SU(3)$ inspired weighting $$\Pi^{3\gamma}_{uds}=\frac12\,
\Pi^{\gamma \gamma}_{uds}$$
the $\rho$ dominance (exact in the isospin limit) assignment reads
$$\Pi^{3\gamma}_{ud}=\frac{1}{2}\, \Pi^{\gamma \gamma}_{ud}\semis
\Pi^{3\gamma}_{s}=\frac{3}{4}\, \Pi^{\gamma \gamma}_{s}$$ which agrees
well with lattice data.

On the $\epm$ data side, I apply flavor separation by hand,
in particular for extracting the isovector part:
we skip all final states involving photons like: $\pi^0\gamma$, $\eta
\gamma$ channels,

as $ud,\;I=0$ we include states with odd number of pions

as $ud,\;I=1$ we include states with even number of pions

as $\bar{s}s$ we count all states with Kaons

States $\eta X$ with $X$ some other hadrons are collected separately, and
then split into $q=u,d$ and $s$ components by experimentally established mixing.

Flavor separation is possible only in regions where exclusive channel
cross sections are available. We perform this in the region 0.61 GeV
to 2.1 GeV. Above this energy  only inclusive $R(s)$ measurements are available,
and a pQCD reweighting is applied.

\begin{figure}[h]
\centering
\sidecaption
\includegraphics[width=0.46\textwidth,clip]{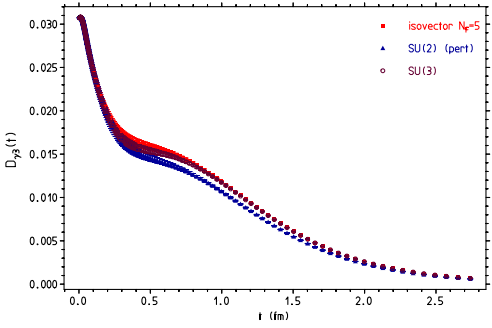}
\caption{$D_{\gamma3}(t)$ versions of flavor separation   a) VMD
isovector, b) in the $\SU(2)$ and neglecting OZI suppressed terms =
perturbative reweighting, with c) flavor separation in the $\SU(3)$ limit
including OZI suppressed contributions.  Version a) fits best to
lattice data, c) shows also reasonable agreement, while b) is
significantly off, i.e. perturbative reweighting and/or neglecting OZI
suppressed effects is inadequate.}
\label{fig:3Gvers}
\end{figure}

\begin{figure}[h]
\centering
\includegraphics[width=0.46\textwidth]{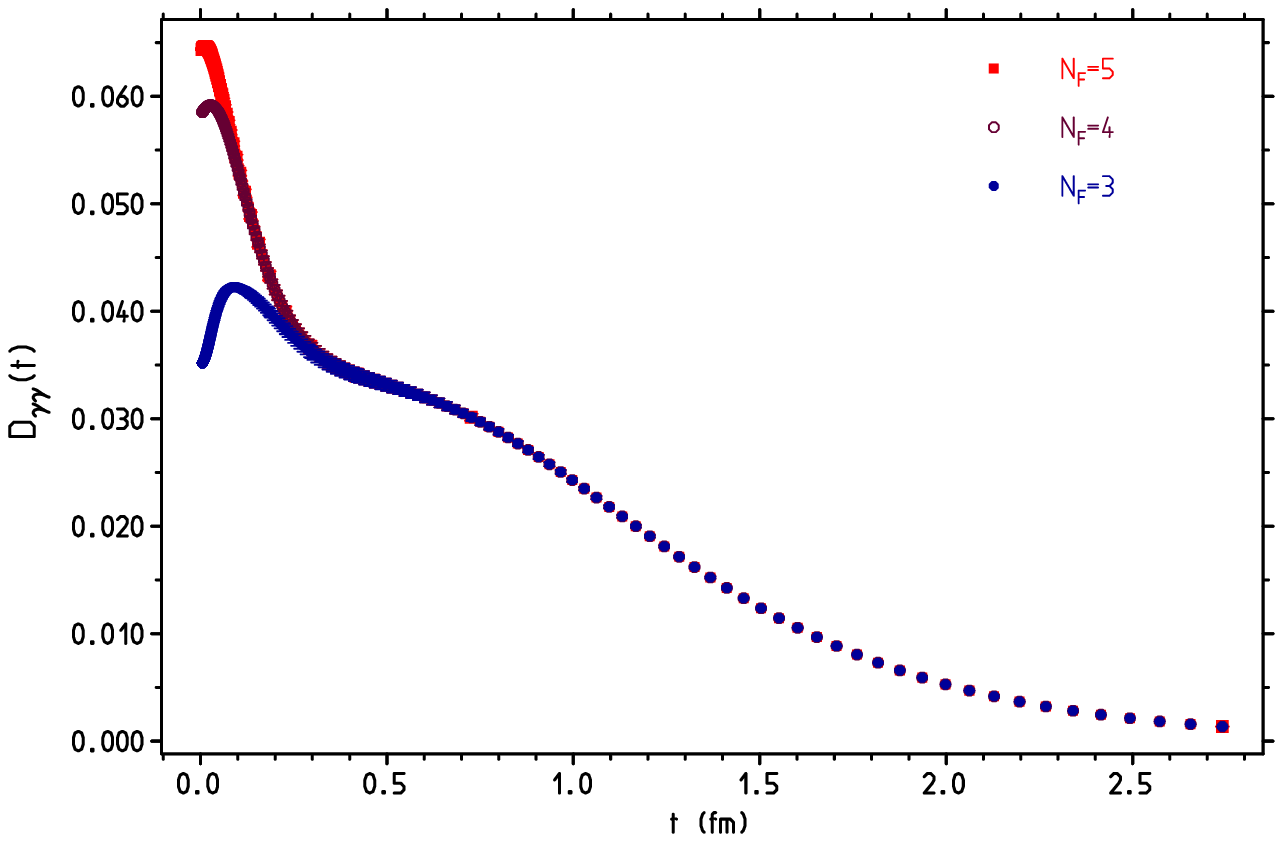}
\includegraphics[width=0.46\textwidth]{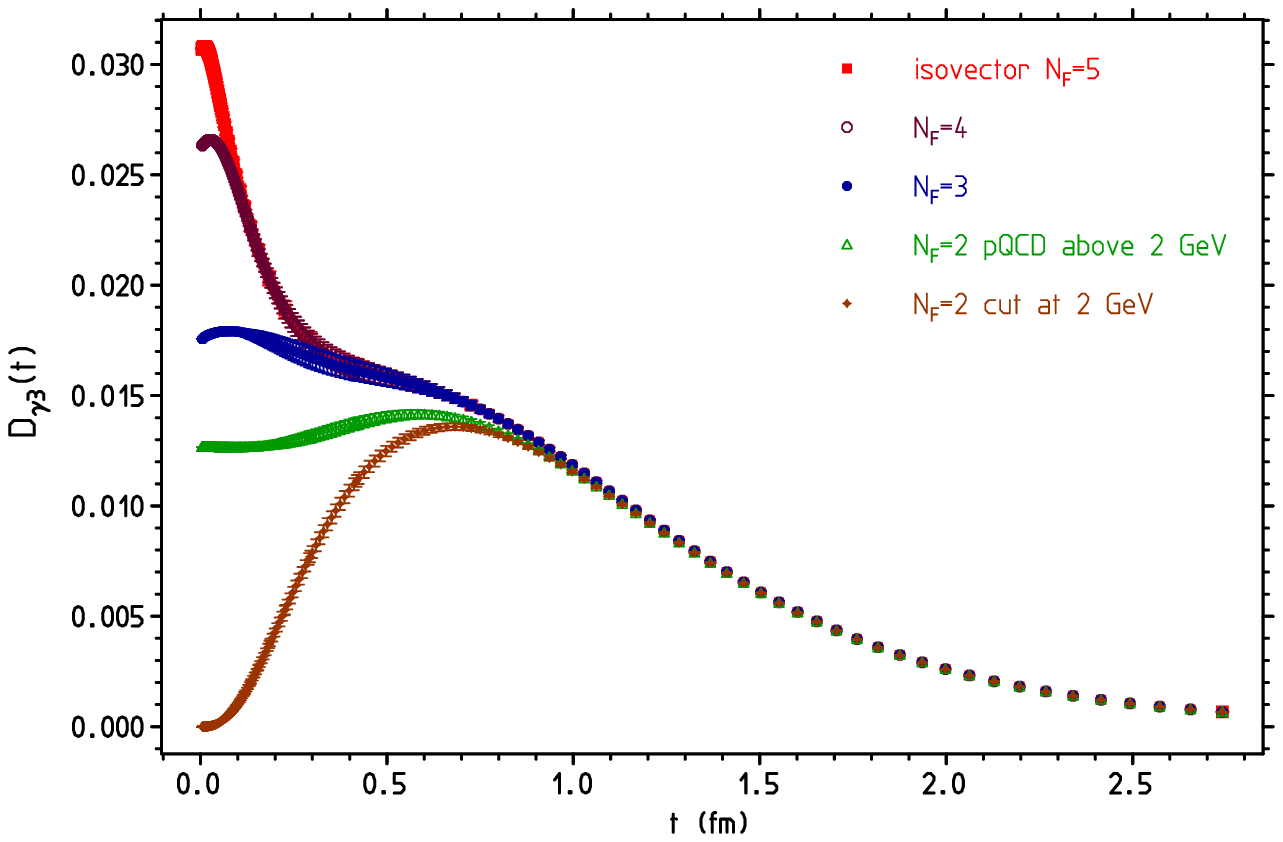}
\caption{Euclidean correlators. Left: $\langle \gamma \gamma \rangle$ for
$N_f=3,4$ and $5$ flavors. Right: the same for the $\langle 3 \gamma \rangle$ time correlator
In the ($u,d$) sector $\langle 3
\gamma \rangle$ is $1/2\: \langle \gamma \gamma \rangle$.}
\label{fig:ECGG345}
\end{figure}

\begin{table}
\centering
\caption{Variants of flavor recombination of $\braket{3\gamma}$ in
terms of $\braket{\gamma \gamma}$. LQCD tests strongly disfavor
``$\SU(2)$''~\cite{Francis:2013jfa,Burger:2015lqa}}
\label{tab:flavor}
\begin{tabular}{ccrcrlcc}
\hline\noalign{\smallskip}
variant && \multicolumn{3}{c}{weights} & ``model'' & {\tt alphaQED}&\\
\noalign{\smallskip}\hline\noalign{\smallskip}
 $\SU(3)$ & = & $\frac12\,[ud]^{I=1}$ & + & $\frac12 [s]$ & assuming
 $\SU(3)$ symmetry & {\tt hadr5n09} &\\[1mm]
 ``$\SU(2)$'' & = & $\frac{9}{20}\,[ud]^{I=1}$& +&$ \frac34 [s]$ &
 perturbative reweighting & {\tt hadr5n12} &\labelitemx\\[1mm]
 VMD [iso] & = & $\frac12\,[ud]^{I=1}$& + &$\frac34 [s]$ & VMD
 isovector & {\tt hadr5n16/17} &\labelitemv\\
\noalign{\smallskip}\hline
\end{tabular}
\end{table}
Key objects in lattice QCD are Euclidean time correlators:
\be
I(t)=t^3 \int_a^\infty \D \,\omega\, \omega^2\, \rho (\omega^2)\,
\E^{-\omega t}\semis \rho(s)=\frac{R(s)}{12\pi^2}\epo
\ee
Normalization (as in ~\cite{Jegerlehner:1985gq} i.e. as weak currents in SM):
$D_{\gamma \gamma}(t)=\braket{\gamma \gamma}(t)\semis
D_{\gamma 3}(t)=\frac12\,\braket{\gamma 3}(t)\epo$ 
The Euclidean time variable $t$ is in units of 1 fermi fm = 0.1973269631
in GeV$^{-1}$, i.e. $t=\mathrm{fm}/E[\mathrm{GeV}]$.
For $R(s)=1$ the integral is given by
\bea
I(t,a,L)[R=1]= \frac{1}{12\pi^2}\,t^3 \int_a^L \D \,\omega\, \omega^2\,
\E^{-\omega t} =\frac{1}{12\pi^2}\,\left\{
\left(a^2t^2+2at+2\right)\,\E^{-at}-\left(L^2t^2+2 Lt+2\right)\,\E^{-Lt}\right\}\epo
\eea
Calculated in terms of $R(s)$ the flavor-recombination variants listed
in Table~\ref{tab:flavor} are compared in Fig.~\ref{fig:3Gvers} and results
for the ``best fit'' are shown in Fig.~\ref{fig:ECGG345} for different
flavor contents\footnote{One may
calculate $\amu$ directly in terms of the Euclidean time correlator
as~\cite{Taku}
\ba
a_\mu^{\rm HVP-LO}=4\,\alpha^2\,m_\mu \int\limits_0^{\infty} \D t \,t^3\,I(t)\,\overline{K}(t)
\ea
with kernel
\ba
\overline{K}(t)=\frac{2}{m_\mu\,t^3}\int\limits_0^{\infty} \frac{\D Q}{Q}\,
f(Q^2)\,\left[(Q/E_0)^2-4\,\sin^2 \left(\frac12\,Q/E_0\right) \right]_{E_0=1/t}
\ea
and
\ba
f(s)=\frac{1}{m_\mu^2}\,rZ(r)^3\,\frac{1-rZ(r)}{1+rZ(r)^2}\semis
Z(r)=\frac{\left(\sqrt{r^2+4r}-r\right)}{2 r}\semis
\semis r=s/m_\mu^2\epo
\ea
I find \mbo{\amuh=685.58 (1.30)(4.85)\power{-10}}
from Euclidean time correlator
for the HVP LO contribution (obtained as a 2-step integration), in very good
agreement with the result of the direct integration of
$R(s)$ \mbo{\amuh=686.04 (0.90)(4.09)\power{-10}}.}.

\section{IB and EM corrections to lattice QCD HVP results}
Lattice QCD ab initio calculations of Euclidean current correlators
come closer to produce results providing important crosschecks of the
standard dispersion relation approach based on $\epm$ data. Here, in
Table~\ref{tab:amIBEMNC} I provide and update for $\amu$, $a_e$ and
$a_\tau$, respectively, isospin breaking (IB) and electromagnetic (EM)
corrections not included so far in lattice calculations. A detailed
description of the calculations may be found in my
book~\cite{Jegerlehner:2017gek}.  After submitting the manuscript of
the book, I had more time to think carefully about the isospin and
e.m. corrections. So I found one of the corrections concerning the
dependence on the pion mass not to be the relevant answer to the
question what would be the change of a $m_{\pi^0} \to m_{\pi^\pm}$
shift in lattice results. The shift has been estimated using the
Gounaris-Sakurai (GS) parametrization\footnote{Specifically, I use GS
neutral channel (NC)~\cite{Akhmetshin:2001ig} Eqs. 8 to 18 and GS
charged channel (CC) see~\cite{Fujikawa:2008ma} Eqs. 11 to 16 }, which
however \textbf{has not the correct dependence on the pion mass},
because it includes $M_\rho$, $\Gamma_\rho$ and $m_{\pi}$ as
independent parameters and the shift has been calculated at fixed
resonance mass and width. Changing $m_\pi$ in the standard
Gounaris-Sakurai parametrization (as commonly done in calculating IB
effects for the relation between CC (tau) and NC (ee) channels), this
is only a partial effect, as the GS formula includes the pion mass
dependence in some hidden form. When one uses instead a QFT version as
discussed e.g. in~\cite{Jegerlehner:2011ti} (i.e. a QFT provided form of the
Breit-Wigner) or also as modeled by he HLS approach one obtains a
very different pion mass dependence, as given now in a modified
table. The pion mass shift in $|F_\pi|^2$ is now much larger and
compensates largely the large shift in the relation between $R(s)$ and
$|F_\pi|^2$.

So there is an update of Table 5.24 of the
book~\cite{Jegerlehner:2017gek} (entries concerning the pion mass
dependence) to be replaced by Table~\ref{tab:amIBEMNC}.

%
%
%
%
%
%

My suspicion that something must be wrong with the GS estimate of the
pion mass shift I had when I looked at the shift in the width of the
$\rho$ from $m_\pi \to m_{\pi^0}$, which is actually large (about 2 MeV)
but seemed to have a small effect on $|F_\pi|^2$, which turns out to be
an outcome of the GS form.

If one considers the QFT version of the Breit-Wigner, one can see that the
cross section $\sigma_{\rm BW}$ at peak
\bea
\sigma_{\rm BW} = \frac{12 \pi}{M_\rho^2}\,\frac{\Gamma_{ee}}{\Gamma_{\rm tot}} \mathrm{ \ \ at \
peak}
\eea
only depends on the $\rho$ mass and the ratio
$\Gamma_{ee}/\Gamma_{\rm tot}$ at $M_\rho$, so the dependence on
$m_\pi$ must be small\footnote{The velocity factors 
which cause the large shifts in the widths are common in $\Gamma_{ee}$ and
$\Gamma_{\rm tot}$ and thus drop out in the cross-section. The
parameter to be kept fixed if the dimensionless $\rho \to \pi\pi$
coupling $g_{\rho\pi\pi}$. } and results from the fact the the $\pi\pi$
channel is not 100\% saturated by the $\rho$ meson.  I advocate to
perform the \mbo{m_{\pi^0} \to m_{\pi^\pm}} extrapolation on lattice
data directly! Otherwise, utilizing a GS ansatz for the extrapolation
of lattice data in the pion mass, requires to take into account the
proper pion mass dependence of mass and width of the vector resonance
as well.

One is always tempted to take the GS parametrization of the $\pi\pi$
data because experiments as well as the PDG still are extracting the
$\rho$ parameters by using the GS formula, which we criticized
in~\cite{Jegerlehner:2011ti}. The VMD I ansatz on which GS is based
has actually has been criticized by Kroll, Lee and Zumino in 1967
already for lack of e.m. gauge invariance.

{\small
\begin{table}[h!]
\centering
\caption{Neutral channel: muon, electron and $\tau$ missing effects in lattice QCD simulations performed in the
isospin limit $m_d=m_u$ and without QED effects. Effects have been
integrated from 300~MeV to 1~GeV}
\label{tab:amIBEMNC}
\begin{tabular}{lcccccc}
\hline\noalign{\smallskip}
& \multicolumn{2}{c}{$\delta a_\mu\power{10}$}
& \multicolumn{2}{c}{$\delta a_e\power{14}$}
& \multicolumn{2}{c}{$\delta a_\tau\power{8}$} \\
\hline\noalign{\smallskip}
Correction type & GS fit & shift &GS fit & shift &GS fit & shift \\
\noalign{\smallskip}\hline\noalign{\smallskip}
$I=1$ NC: GS fit of $\epm$ data~$^{[1]}$ &
489.21$^\star$ & & 
134.49$^\star$ & &
167.66$^\star$ & \\
$\omega-\rho$ mixing & 
491.89 & +2.68 &
135.24 & +0.75 &
168.39 & +0.73 \\
FSR of $ee$ $I=1+0$ & 
496.11 & +4.22 &
136.41 & +1.17 &
169.80 & +1.41 \\
$\gamma-\rho$ mixing & 
486.47 & -2.74 &
133.99 & -0.50 &
165.14 & -2.52\\
\noalign{\smallskip}\hline
Elmag. shift $m_{\pi^0} \to m_{\pi^\pm}$ & & shift of $^\star$ &&\\
\noalign{\smallskip}\hline
$I=1$ NC $m_\pi=m_{\pi^0}$ $R(s)$ vs. $|F_\pi|^2$~$^{[2]}$&
502.01 & +12.81&
138.21 & +3.72&
171.22 & +3.56\\
$I=1$ NC $m_\pi=m_{\pi^\pm}$ $|F_\pi|^2$~$^{[3]}$& 455.89 & 
& 125.76 & 
& 154.23 &  \\
$I=1$ NC $m_\pi=m_{\pi^0}$ $|F_\pi|^2$ & 
441.97 & -13.92 &
121.85 & -3.91 &
150.05 & -4.18\\
\noalign{\smallskip}\hline
Combined $m_\pi=m_{\pi^0}$ & 
500.91 & &
137.91 & &
170.83 & \\
Physical $m_\pi=m_{\pi^\pm}$~$^{[4]}$& 
489.20 & 1.12&
134.49 & 0.19&
167.66 & 0.62\\
\noalign{\smallskip}\hline
Elmag. channels~\cite{HLS12} & & & & && \\
\noalign{\smallskip}\hline
$\pi^0 \gamma$ & 
\multicolumn{2}{c}{$4.64\pm 0.04$}& 
\multicolumn{2}{c}{$1.33\pm 0.04$}&
\multicolumn{2}{c}{$2.11\pm 0.05$}\\ 
$\eta \gamma$ &
\multicolumn{2}{c}{$0.65\pm 0.01$}&
\multicolumn{2}{c}{$0.17\pm 0.00$}&
\multicolumn{2}{c}{$0.33\pm 0.01$}\\
$\pi^+\pi^-\pi^0$
missing  disconnected ?& 
\multicolumn{2}{c}{$5.26\pm 0.15$}&
\multicolumn{2}{c}{$1.76\pm 0.06$}&
\multicolumn{2}{c}{$2.90\pm 0.10$}\\
\noalign{\smallskip}\hline
\end{tabular}

$^{[1]}$ $\omega$ switched off,~~
$^{[2]}$ [$|F_\pi|^2$ fixed],~~
$^{[3]}$ [BW $\rho$ FF],~~
$^{[4]}$ plus e.m. shift in mass\&width of the $\rho$\hfill
\end{table}}
For the charged channel the corresponding results are collected in Table~\ref{tab:amIBEMCC}.
\begin{table}
\centering
\caption{Charged channel: missing effects in lattice QCD simulations performed in the
isospin limit $m_d=m_u$ and without QED effects. Tabulated are the
effects $\delta a_\ell$ ($\ell=\mu.e,\tau$) integrated from 300~MeV to 1~GeV}
\label{tab:amIBEMCC}
\begin{tabular}{llclclc}
\hline\noalign{\smallskip}
& \multicolumn{2}{c}{$\delta a_\mu\power{10}$}
& \multicolumn{2}{c}{$\delta a_e\power{14}$}
& \multicolumn{2}{c}{$\delta a_\tau\power{8}$} \\
\hline\noalign{\smallskip}
Correction type & GS fit & shift& GS fit & shift& GS fit & shift  \\
\noalign{\smallskip}\hline\noalign{\smallskip}
GS fit of $\tau$ data                  & 505.32 &        
&  139.22 &        
&  171.35 &         \\
$+\delta M_\rho,\,+\delta \Gamma_\rho$ & 501.44 & -3.88  
&  138.16 & -1.06  
&  170.04 & - 1.31  \\
$1/G_{\rm EM}$                         & 504.62 & -0.70  
&  138.94 & -0.28  
&  171.51 & + 0.16  \\
$\beta_-^3/\beta_0^3$                  & 498.73 & -6.59 
&  137.30 & -1.92  
&  169.53 & - 1.82  \\
\noalign{\smallskip}\hline
$I=1$, LQCD type                       & 494.15 & -11.17 
& 135.96 & -3.26   
&  168.38 & - 2.97 \\
\noalign{\smallskip}\hline
\end{tabular}
\end{table}
Summing up the various corrections yields the results listed in Table~\ref{tab:IBEMfin}. 
\begin{table}
\centering
\caption{Neutral channel: total shifts for \mbo{a_\ell} ($\ell=\mu,e,\tau$)}
\label{tab:IBEMfin}
\begin{tabular}{lrrr}
\hline\noalign{\smallskip}
type of correction & \mbo{\delta a_\mu\power{10}} & \mbo{\delta a_e
\power{12}} & \mbo{\delta a_\tau \power{8} }\\
\noalign{\smallskip}\hline\noalign{\smallskip}
  iso+em from \mbo{\pi\pi} channel :& +4.16(4) & + 1.42(1) & -0.38(0)\\
 incl e.m. decays \mbo{\pi^0\gamma} and \mbo{\eta\gamma}:& + 5.29(4) &  + 1.19(4) & + 2.06(7) \\
 missing \mbo{\phi \to \pi^+\pi^-\pi^0} ?:
 & + 5.26(15) & + 1.35(4) & + 2.78(8) \\
sum & 14.71(16)    & 3.96(6) &  4.46(11)\\
\noalign{\smallskip}\hline
\end{tabular}
\end{table}

Which of the corrections has to be supplemented depends on the what and
whatnot has been included in a given lattice QCD calculation.

\clearpage

\textbf{Acknowledgments:}
I thank the organizers of the Phi to Psi 2017 Workshop at
Mainz for the kind invitation and the kind hospitality and DESY for the support.


\begin{thebibliography}{99}
\small
%
\bibitem{Jegerlehner:2017gek}
  F.~Jegerlehner,
  \textit{The Anomalous Magnetic Moment of the Muon},
  Springer Tracts Mod.\ Phys.\  {\bf 274}, pp.1 (2017), doi:10.1007/978-3-319-63577-4
%
\bibitem{Jegerlehner:2017lbd}
  F.~Jegerlehner,
  arXiv:1705.00263 [hep-ph].

\bibitem{Jegerlehner:2015stw}
  F.~Jegerlehner,
  EPJ Web Conf.\  {\bf 118}, 01016 (2016)

\bibitem{LeeRoberts}
B.~Lee Roberts, \textit{FNAL $(g-2)_\mu$ Experiment}, these proceedings

\bibitem{Mibe}
Tsutomu Mibe, \textit{JPARC $(g-2)_\mu$ Experiment}, these proceedings

\bibitem{Colangelo}
Gilberto Colangelo, \textit{HLBL Dispersive theory Bern}, these proceedings

\bibitem{Pauk}
Vladiszlav Pauk, \textit{HLBL Dispersive theory Mainz}, these proceedings

\bibitem{Kurz:2014wya}
  A.~Kurz, T.~Liu, P.~Marquard, M.~Steinhauser,
  Phys.\ Lett.\ B {\bf 734}, 144 (2014)

\bibitem{Zhang:2015yfi}
  Z.~Zhang,
  EPJ Web Conf.\  {\bf 118}, 01036 (2016) and these proceedings

\bibitem{Hagiwara:2017zod} 
  K.~Hagiwara {et al.},
  Nucl.\ Part.\ Phys.\ Proc.\  {\bf 287-288}, 33 (2017);
 T.~Teubner, these proceedings

\bibitem{Aoyama:2012wj}
  T.~Aoyama, M.~Hayakawa, T.~Kinoshita, M.~Nio,
  Phys.\ Rev.\ Lett.\  {\bf 109}, 111807 (2012);
%
ibid. 111808 (2012);
%
  Phys.\ Rev.\ D {\bf 91}, 033006 (2015)

\bibitem{Laporta:2017okg}
  S.~Laporta,
  Phys.\ Lett.\ B {\bf 772}, 232 (2017)

\bibitem{Steinhauser}
  A.~Kurz et al.,
  PoS LL {\bf 2016}, 009 (2016);
M.~Steinhauser, these proceedings

\bibitem{Gnendiger:2013pva}
  C.~Gnendiger, D.~St\"ockinger, H.~St\"ockinger-Kim,
  Phys.\ Rev.\ D {\bf 88}, 053005 (2013)

\bibitem{BNL04} G.~W.~Bennett et al. [Muon (g-2) Collab.],
Phys.\ Rev.\ Lett.\  \textbf{92}, 161802 (2004)

\bibitem{Krauth:2017ijq} 
  J.~J.~Krauth et al.,
  arXiv:1706.00696 [physics.atom-ph].

\bibitem{PRP}
A.~Beyer et al.,
Science, {\bf 358}:79 (2017), DOI: 10.1126/science.aah6677.

\bibitem{aenew}
B.~Odom, D.~Hanneke, B.~D'Urso, G.~Gabrielse
Phys.\ Rev.\ Lett.\  {\bf 97}, 030801 (2006)

\bibitem{EJKV98}
S.~Eidelman, F.~Jegerlehner, A.~L.~Kataev, O.~Veretin,
Phys.\ Lett.\ B {\bf 454}, 369  (1999)

\bibitem{FJ98} F. Jegerlehner, In: {\it Radiative Corrections},
ed J.~Sol\`a (World Scientific, Singapore 1999) pp 75--89

\bibitem{Jegerlehner:2008rs} 
  F.~Jegerlehner,
  Nucl.\ Phys.\ Proc.\ Suppl.\  {\bf 181-182}, 135 (2008)

\bibitem{Chetyrkin:1996cf} 
  K.~G.~Chetyrkin, J.~H.~K\"uhn, M.~Steinhauser,
  Nucl.\ Phys.\ B {\bf 482}, 213 (1996)

\bibitem{Chetyrkin:1997qi} 
  K.~G.~Chetyrkin, R.~Harlander, J.~H.~K\"uhn. M.~Steinhauser,
  Nucl.\ Phys.\ B {\bf 503}, 339 (1997)

\bibitem{Davier:2003pw} 
  M.~Davier, S.~Eidelman, A.~H\"ocker, Z.~Zhang,
  Eur.\ Phys.\ J.\ C {\bf 31}, 503 (2003)

\bibitem{Ghozzi:2003yn} 
  S.~Ghozzi, F.~Jegerlehner,
  Phys.\ Lett.\ B {\bf 583}, 222 (2004)

\bibitem{Davier:2009ag} 
  M.~Davier {et al.},
  Eur.\ Phys.\ J.\ C {\bf 66}, 127 (2010)

\bibitem{KLOE-2:2016mgi}
  A.~Anastasi et al. [KLOE-2 Collab.],
  Phys.\ Lett.\ B {\bf 767}, 485 (2017);\\
  G.~Venanzoni [KLOE-2 Collab.],
  arXiv:1705.10365 [hep-ex] and these proceedings

\bibitem{Jegerlehner:1985gq}
  F.~Jegerlehner,
  Z.\ Phys.\ C {\bf 32}, 195 (1986)

\bibitem{Jegerlehner:2011mw}
  F.~Jegerlehner,
  Nuovo Cim.\  {\bf 034C}, 31 (2011)
  [arXiv:1107.4683 [hep-ph]]

\bibitem{alphaQED17}
{\tt {http://www-com.physik.hu-berlin.de/\mytexttilde{}fjeger/alphaQED17.tar.gz}}

\bibitem{Meyer:2011um}
  H.~B.~Meyer,
  Phys.\ Rev.\ Lett.\  {\bf 107}, 072002 (2011)

\bibitem{Bernecker:2011gh}
  D.~Bernecker, H.~B.~Meyer,
  Eur.\ Phys.\ J.\ A {\bf 47}, 148 (2011)

\bibitem{Francis:2013jfa}
  A.~Francis, G.~von Hippel, H.~B.~Meyer, F.~Jegerlehner,
  PoS LATTICE {\bf 2013}, 320 (2013)

\bibitem{Burger:2015lqa}
  F.~Burger, K.~Jansen, M.~Petschlies, G.~Pientka,
  JHEP {\bf 1511}, 215 (2015)

\bibitem{Akhmetshin:2001ig} 
  R.~R.~Akhmetshin {et al.} [CMD-2 Collaboration],
  Phys.\ Lett.\ B {\bf 527}, 161 (2002)

\bibitem{Fujikawa:2008ma} 
  M.~Fujikawa {et al.} [Belle Collaboration],
  Phys.\ Rev.\ D {\bf 78}, 072006 (2008)

\bibitem{Jegerlehner:2011ti} 
  F.~Jegerlehner, R.~Szafron,
  Eur.\ Phys.\ J.\ C {\bf 71}, 1632 (2011)

\bibitem{Taku}
T.~Izubuchi, private communication

\bibitem{HLS12}
  M.~Benayoun, P.~David, L.~DelBuono, F.~Jegerlehner,
  Eur.\ Phys.\ J.\ C {\bf 72} (2012) 1848

\end{thebibliography}
\end{document}